\documentclass[prb,aps,showpacs,superscriptaddress, floatfix]{revtex4}
\usepackage{graphicx}
\usepackage{bm}
\usepackage{amsmath}
\begin{document}
\title{
Half-skyrmion picture of single hole doped CuO$_2$ plane
}
\author{Takao Morinari}
\affiliation{Yukawa Institute for Theoretical Physics, Kyoto
University, Kyoto 606-8502, Japan}
\date{\today}
\begin{abstract}
Based on the Zhang-Rice singlet picture, it is argued that the
half-skyrmion is created by the doped hole in the single hole doped
high-$T_c$ cuprates with N{\' e}el ordering.
The spin configuration around the Zhang-Rice singlet,
which has the form of superposition of the two different d-orbital
hole spin states, is studied within 
the non-linear $\sigma$ model and the CP$^1$ model.
The spin configurations associated with each hole spin state are
obtained, and we find that the superposition of these spin
configuration turns out to be the half-skyrmion that is
characterized by a half of the topological charge.
The excitation spectrum of the half-skyrmion is obtained by making use 
of Lorentz invariance of the effective theory and is qualitatively in
good agreement with angle resolved photoemission spectroscopy on the
parent compunds.
Estimated values of the parameters contained in the excitation
spectrum are in good agreement with experimentally obtained values.
The half-skyrmion theory suggests a picture for the
difference between the hole doped compounds and the electron doped
compounds.
\end{abstract}



\maketitle

\section{Introduction}
In the phase diagram of high-temperature superconductors,
there are a variety of phases:
the N{\' e}el ordering phase, the spin-glass like phase, the d-wave
superconducting phase, and the pseudogap phase.\cite{TIMUSK_STATT}
This rich phase diagram is controlled by the hole doping concentration
$x$ and temperature.
If we focus on the ground state properties, then the fundametal
parameter is $x$.
In order to understand the physics of high temperature
superconductivity, it is necessary to figure out how to describe
the doped holes.
The goal of this approach is, of course, to find a unfied description
of the doped holes over the whole range of $x$.

In this paper, we consider the simplest case:
We focus on the single hole doped system.
The motivation is the following.
Although d-wave superconductivity occurs
in moderately doping region, and there is the intriguing pseudogap
phase, 
it is quite hard to find a reliable description of the doped holes
because of strong correlation effects between the holes and background
spin fluctuations.
In contrast, the physics of the undoped compound is well
established compared with other phases.
The system is described by the $S=1/2$ antiferromagnetic Heisenberg
model on the square lattice and the ground state is the N{\' e}el ordered
state.
Considering one hole doping upon this well-established phase would be
the first step to understand the effects of doped holes in the
high-temperature superconductors.

Experimentally it has been known that doped holes occupy oxygen
p-orbitals in the CuO$_2$ plane.
Between an oxygen p-orbital hole and the nearest neighbor 
copper d-orbitals holes, there is strong correlations of forming a
singlet pair.
The resulting singlet state is called Zhang-Rice
singlet.\cite{ZHANG_RICE}
Based on this picture, the t-J model was proposed.

In terms of the t-J model, the single hole problem has been discussed
extensively.\cite{KANE_ETAL}
However, the focus is mainly on frustration effect induced by hopping 
of the doped hole.
Not so much attention has been paid on the spin configuration around
the Zhang-Rice singlet state.
In this paper, we study the spin configuration created around the
Zhang-Rice singlet within effective field theory approaches to the
Heisenberg antiferromagnet.

As an effective theory for the $S=1/2$ Heisenberg antiferromagnet on
the square lattice, the non-linear $\sigma$ model (NL$\sigma$M) has
been studied extensively:\cite{CHN}
\begin{equation}
S=\frac{\rho_s}{2} \int_0^{\beta} d\tau \int d^2 {\bf r}
\left[ \left( \nabla {\bf n} \right)^2
+ \frac{1}{c_{\rm sw}^2} 
\left( \frac{\partial {\bf n}}{\partial \tau} \right)^2
\right],
\label{eq_nlsm}
\end{equation}
where $\rho_s$ is the spin stiffness and $c_{\rm sw}$ is the
antiferromagnetic spin wave velocity.
(Hereafter we take the unit of $\hbar =1$.)
Here $\tau$ is the imaginary time and $\beta = (k_B T)^{-1}$ with $T$
being temperature.
The unit vector ${\bf n}$ represents the staggered momemnt.
This model is derived from the Heisenberg antiferromagnet by applying
Haldane's mapping.\cite{Haldane}
Theoretical formula for the antiferromagnetic correlation length
$\xi_{AF}$ based on the renormalization group analysis of this model
is in quite good agreement with experimentally obtained $\xi_{AF}$.

Another well-known approach is the Schwinger boson mean field
theory (SBMFT).\cite{AROVAS_AUERBACH}
In this theory, the spin $S=1/2$ is represented by 
\begin{equation}
{\bf S}_j = \frac12 
\left(
\begin{array}{cc}
z_{j\uparrow}^{\dagger} & z_{j\downarrow}^{\dagger}
\end{array} 
\right)
{\mbox{\boldmath ${\bf \sigma}$}}
\left(
\begin{array}{c}
z_{j\uparrow}\\
z_{j\downarrow}
\end{array} 
\right),
\label{eq_SBS}
\end{equation}
with the constraint $\sum_{\sigma} z_{j\sigma}^{\dagger} z_{j\sigma}
=1$.
The components of ${\mbox{\boldmath ${\bf \sigma}$}}$ are Pauli
matrices.
The antiferromagnetic correlations are described by a mean field
$A_{ij}=\langle z_{i\uparrow} z_{j\downarrow}
-z_{i\downarrow} z_{j\uparrow} \rangle$.
This mean field describes also the pairing correlations of the
Schwinger bosons.\cite{READ_SACHDEV,CHUBUKOV,NG95}
The quasiparticle excitation spectrum of the Schwinger bosons is given
by
\begin{equation}
\omega_k = \sqrt{\lambda^2 - 4J^2 A^2 \gamma_k^2},
\end{equation}
where $\gamma_k = (\sin k_x \pm \sin k_y)/2$ with
the plus sign is for $k_x k_y>0$ and
the minus sign is for $k_x k_y <0$.
The parameter $\lambda$ is originally introduced as a Lagrange
multiplier to impose the constraint.
In the mean field approximation, the Lagrange multiplier is taken to
be uniform.
For the ground state, $\lambda=2JA$.
Bose-Einstein condensation of the Schwinger bosons at ${\bf
k}=(\pm \pi/2,\pm \pi/2)$ leads to N{\' e}el ordering.\cite{BEC}

In order to consider fluctuations about Bose-Einstein condensate of
Schwinger bosons, it is convenient to use a gauge field description.
Introducing a variable $x_0 = c_{\rm sw} \tau$, the action reads
$S=\frac{1}{2g} \int d^3 x \left( \partial_{\mu} {\bf n} \right)^2$
with $g=c_{\rm sw}/\rho_s$.
In terms of complex fields $\overline{\zeta}_{\sigma}$ and
$\zeta_{\sigma}$
with $\sum_{\sigma} \overline{\zeta}_{\sigma} \zeta_{\sigma}=1$, the
vector ${\bf n}$ is represented by ${\bf n} = \overline{\zeta} 
{\mbox{\boldmath ${\bf \sigma}$}} \zeta$.
Substituting this into the action, we obtain
\begin{equation}
S=\frac{2}{g} \int d^3 x \left[
(\partial_{\mu} \overline{\zeta} )
(\partial_{\mu} \zeta ) 
+ \left( \overline{\zeta} \partial_{\mu} \zeta \right)^2 \right].
\label{eq_cp1p}
\end{equation}
Performing a Stratonovich-Hubbard transformation at the interaction
term, we obtain
\begin{equation}
S=\frac{2}{g} \int d^3 x \sum_{\sigma}
\left| \left( \partial_{\mu} -i \alpha_{\mu} \right)
\zeta_{\sigma} (x) \right|^2.
\label{eq_cp1}
\end{equation}
Note that (\ref{eq_cp1p}) is invariant under a local U(1) gauge
transformation $\zeta\rightarrow \zeta e^{i\theta(x)}$ 
and $\overline{\zeta}\rightarrow \overline{\zeta} e^{-i\theta(x)}$.
The field $\alpha_{\mu}$ is a gauge field assocaited with this gauge
invariance.
As explicitly shown in Ref.\cite{READ_SACHDEV}, the CP$^1$ model is
also derived from SBMFT.
The relation between the fields $\zeta$ and $\overline{\zeta}$ and the 
Schwinger boson fields (\ref{eq_SBS}) is given by
$\zeta_{j\sigma} = z_{j\sigma}$ for one sublattice and
$\zeta_{j\sigma} = \overline{z}_{j\sigma}$ for the other sublattice.

In addition to these bosonic models, there is a fermionic theory.
For $S=1/2$ spins, one can represent them by fermions,
\begin{equation}
{\bf S}_j = \frac12 
\left(
\begin{array}{cc}
f_{j\uparrow}^{\dagger} & f_{j\downarrow}^{\dagger}
\end{array} 
\right)
{\mbox{\boldmath ${\bf \sigma}$}}
\left(
\begin{array}{c}
f_{j\uparrow}\\
f_{j\downarrow}
\end{array} 
\right),
\end{equation}
with the constraint 
$\sum_{\sigma} f_{j\sigma}^{\dagger} f_{j\sigma}=1$.
Based on a mean field theory, the $\pi$-flux phase was proposed by
Affleck and Marston.\cite{AFFLECK_MARSTON}
The excitation spectrum of the quasiparticles in the $\pi$-flux phase
is given by
\begin{equation}
\epsilon_k = \pm v\sqrt{\cos^2 k_x + \cos^2 k_y},
\end{equation}
where $v$ is a mean field paramter of the theory.
Fluctuations about this mean field state can be taken into account
through a gauge field.
For the two-dimensional $S=1/2$ quantum Heisenberg antiferromagnet,
the condition for the dynamical mass generation is
satisfied,\cite{MARSTON,KIM_LEE} and so
the spectrum is modified as
\begin{equation}
\epsilon_k = \pm \sqrt{v^2 (\cos^2 k_x + \cos^2 k_y) + m^2}.
\end{equation}
The mass $m$ is associated with the ordered staggered
moment.\cite{KIM_LEE}

For the description of the doped hole, we shall assume that the hole
forms a Zhang-Rice singlet with a copper hole which 
was proposed by Zhang and Rice\cite{ZHANG_RICE} 
from the analysis of the d-p model,
\begin{eqnarray}
H&=&\sum_{j\sigma} \epsilon_d d_{j\sigma}^{\dagger} d_{j\sigma}
+ \sum_{\ell \sigma} \epsilon_p p_{\ell \sigma}^{\dagger} 
p_{\ell \sigma}
+ U \sum_j d_{j\uparrow}^{\dagger} d_{j\uparrow}
d_{j\downarrow}^{\dagger} d_{j\downarrow} \nonumber \\
& & - \sum_{j,\ell,\sigma} \epsilon_{j\ell} t_{pd} 
d_{j\sigma}^{\dagger} p_{\ell \sigma}
+ h.c.,
\end{eqnarray}
where $\epsilon_{j,j-{\hat x}/2}
=\epsilon_{j,j-{\hat y}/2}=-\epsilon_{j,j+{\hat x}/2}=
-\epsilon_{j,j+{\hat y}/2}=1$.
Here $d_{j\sigma}^{\dagger}$ and $p_{\ell\sigma}^{\dagger}$ are the
hole creation operator at the copper d-orbital state and the oxygen
p-orbital state, respectively.
The vacuum is defined as Cu(3d)$^{10}$ and O(2p)$^6$.
Applying a canonical transformation and omitting unimportant terms, 
we obtain
\begin{equation}
H_K = \sum_j 2 \left( \frac{t_{pd}^2}{U-\Delta}
+ \frac{t_{pd}^2}{\Delta} \right)
\left( d_j^{\dagger} {\mbox{\boldmath ${\bf \sigma}$}} d_j \right)
\left( P_j^{\dagger} {\mbox{\boldmath ${\bf \sigma}$}} P_j \right),
\end{equation}
where $P_{j\sigma}=\sum_{\left\{\ell\right\}\in j} 
\epsilon_{j\ell} p_{\ell \sigma}/2$.
Constructing the Wannier wave functions,\cite{ZHANG_RICE} the
Zhang-Rice singlet state is created by the following operator
\begin{equation}
\phi_{{\rm ZR}j}^{\dagger} =\frac{1}{\sqrt{2}}
\left( 
d_{j\uparrow}^{\dagger} 
\phi_{j\downarrow}^{\dagger} 
-
d_{j\downarrow}^{\dagger} 
\phi_{j\uparrow}^{\dagger}\right),
\label{eq_ZRwf}
\end{equation}
where
\begin{equation}
\phi_{j\sigma} = \sum_{j'} 
\left\{  \frac{1}{N} \sum_{\bf k} 
   \frac{\exp \left(i{\bf k}\cdot ({\bf R}_j - {\bf R}_{j'}) \right)}
        {\sqrt{1 -\frac12 \left( \cos k_x + \cos k_y \right) }}
\right\}
P_{j'\sigma}.
\end{equation}

In this paper, we study the spin configuration around a Zhang-Rice
singlet.
We shall show that a spin texture which is called half-skyrmion
characterized by a half of a topological charge is created around the
Zhang-Rice singlet.
The dispersion of the half-skyrmion is given by the same form as that
of the quasiparitcle in the $\pi$-flux phase.
The rest of this paper is organized as follows:
In Sec.\ref{sec_static}, we consider a static Zhang-Rice singlet.
We argue that the spin configuration around such a static Zhang-Rice
singlet is half-skyrmion.
In Sec.\ref{sec_moving}, we construct the moving half-skyrmion state
by making use of Lorentz invariance of the NL$\sigma$M.
The lattice action is derived in Sec.\ref{sec_lattice}.
We shall compare the dispersion of the half-skyrmion with the result
of angle resolved photoemission spectroscopy (ARPES) on the parent
compounds.
The effective theory of the half-skyrmion is obtained by applying
duality mapping in Sec.\ref{sec_duality}.
Section \ref{sec_discussion} devoted to Discussion.
Finally, we summarize the results in Sec.\ref{sec_summary}.

\section{Half-skyrmion solution for static Zhang-Rice singlet}
\label{sec_static}
In this section, we consider the spin configuration induced around a
static Zhang-Rice singlet.
The spin configuration for a moving Zhang-Rice singlet shall be
discussed in Sec.\ref{sec_moving} based on the result of this
section.

The creation operator for a Zhang-Rice singlet residing at the site
$j$ defined by
Eq.~(\ref{eq_ZRwf}) suggests that the spin configuration around a 
static Zhang-Rice singlet is given by superposition of the two
spin configurations:
One is the spin configuration created by a spin up state at the
d-orbital state and the other is the spin configuration created by a
spin down state at the d-orbital state.
We study these states separately, then we consider superposition
of these states.
We assume that the spin state of
$\phi_{j\sigma}$ does not play an important role.
We shall discuss its possible effects in Sec.\ref{sec_discussion}.

In order to study the spin configuration, we use the NL$\sigma$M.
Note that we cannot apply a linear response theory, such as a spin
wave theory, to obtain the spin configuration.
For the d-orbital spin state at the site $j$ which is the same as that 
of the N{\' e}el ordered state of the parent compound before introducing
the hole, the surrounding spins are not so much affected by the fixed
spin state at the site $j$.
However, for the d-orbital spin state which is in the opposite
direction to that of the N{\' e}el ordered state of the parent compound
before introducing the hole, the surrounding spins can change their
directions which are beyond the range of the description of the
linear response theory.

We assume that before introducing the doped hole the spin state at the 
site $j$ is up spin state and the staggered magnetization is in the
$z$ direction.
Under this assumption, the spin up state
does not change the directions of the neighborhood spins.
If we include quantum fluctuation effects, then the staggered moments
of neighboring spins would be enhanced because we fix the state at the 
site $j$ to spin up and quantum fluctuations is quenched at this site.
In other words, the state is in the subspace of eigenstates of the
system with spin up state at the site $j$ d-orbital.
However, for simplicity we do not consider quantum fluctuation effects 
here and we restrict ourselves to a classical spin configuration.
Quantum fluctuation effects shall be taken into account through the
renormalization of the parameters.
Thus, the spin configuration associated with the spin up state
is ${\bf n}({\bf R}_{\ell})=+{\hat e}_z$
for any site $\ell$.

The spin configuration for the spin down state at the site $j$
d-orbital is non-trivial.
We would like to obtain the spin configuration satisfying the 
following boundary conditions:
\begin{equation}
{\bf n}({\bf r}) \rightarrow +{\hat e}_z \hspace{2em}(r\rightarrow
\infty),
\label{eq_b1}
\end{equation}
and
\begin{equation}
{\bf n}({\bf R}_j) = -{\hat e}_z.
\label{eq_b2}
\end{equation}
Since we consider the static spin configuration,
we are interested in the spin configuration that minimizes the energy,
\begin{equation}
E=\frac{\rho_s}{2} \int d^2 {\bf r} (\nabla {\bf n})^2.
\end{equation}
The analysis given below follows a general argument for skyrmion
excitations in the NL$\sigma$M.\cite{RAJARAMAN}
We include the constraint $|{\bf n}|^2=1$ through a Lagrange
multiplier:
\begin{equation}
E=\frac{\rho_s}{2} \int d^2 {\bf r} 
\left[ (\nabla {\bf n})^2
+ \lambda (|{\bf n}|^2-1) \right].
\end{equation}
Taking variation with respect to ${\bf n}$, we obtain
\begin{equation}
\nabla^2 {\bf n} - \lambda {\bf n}=0.
\end{equation}
The Lagrange multiplier $\lambda$ is eliminated by using this equation 
and ${\bf n}^2=1$:
\begin{equation}
\nabla^2 {\bf n} - ({\bf n}\cdot \nabla^2 {\bf n} ){\bf n}=0.
\label{eq_1}
\end{equation}
By using the identity\cite{BELAVIN_POLYAKOV}
\begin{equation}
\int d^2 {\bf r}^2 \left[
\partial_{\mu} {\bf n} \pm \epsilon_{\mu \nu} 
( {\bf n}\times \partial_{\mu} {\bf n} )
\right]^2 \geq 0,
\end{equation}
with $\epsilon_{xx}=\epsilon_{yy}=1$ and
$\epsilon_{xy}=-\epsilon_{yx}=1$,
we find
\begin{equation}
E\geq 4\pi \rho_s |Q|,
\label{eq_lowerb}
\end{equation}
where 
\begin{equation}
Q=\frac{1}{4\pi} \int d^2 {\bf r}~
{\bf n} \cdot (\partial_x {\bf n} \times \partial_y {\bf n}),
\label{eq_Q}
\end{equation}
is called the topological charge.
The equality in Eq.(\ref{eq_lowerb}) is satisfied if and only if
\begin{equation}
\partial_{\mu} {\bf n} \pm \epsilon_{\mu \nu} 
( {\bf n}\times \partial_{\mu} {\bf n} )=0.
\label{eq_2}
\end{equation}

The solution of Eq.(\ref{eq_1}) is divided into sectors with
different $Q$ values.
Since the energy in each sector has the lower bound determined by
Eq.(\ref{eq_lowerb}),
it is enough to solve Eq.(\ref{eq_2}) for our purpose.
To solve this equation, it is convenient to use a variable 
$w=(n_x+in_y)/(1-n_z)$.
In terms of $w$, Eq.(\ref{eq_2}) is
\begin{equation}
\partial_{x} w = -i \partial_{y} w,~
\partial_{x} w = i \partial_{y} w.
\end{equation}
Introducing $z=x+iy$ and $\overline{z}=x-iy$, we find
\begin{equation}
\partial_{z} w = 0,~
\partial_{\overline{z}} w = 0.
\end{equation}
Therefore, the solution satisfies the Cauchy-Rieman eqation.
Analytic function of $z$ or $\overline{z}$ is the solution of
Eq.(\ref{eq_2}).
In terms of $w$ and $\overline{w}$, the vector ${\bf n}$ is described
by
\begin{equation}
{\bf n}=\left( \frac{w+\overline{w}}{|w|^2+1},
-i\frac{w-\overline{w}}{|w|^2+1},
\frac{|w|^2-1}{|w|^2+1}
\right).
\end{equation}
The boundary conditions of (\ref{eq_b1}) and (\ref{eq_b2}) are
\begin{equation}
|w|\rightarrow \infty \hspace{2em}(r\rightarrow \infty),
\end{equation}
and
\begin{equation}
w=0 \hspace{2em}({\bf r}={\bf R}_j),
\end{equation}
respectively.
The solution that satisfies these boundary condition is 
\begin{equation}
w=\frac{z}{\lambda},
\label{eq_sk}
\end{equation}
or
\begin{equation}
w=\frac{\overline{z}}{\lambda},
\label{eq_ask}
\end{equation}
up to a phase factor that is associated with a global rotation of all
spins in the plane.
Here $\lambda$ is a parameter which is associated with the size of the 
spin configuration and ${\bf R}_j$ is taken to be the origin to
simplify the expressions.
The vector ${\bf n}$ representation of Eqs.(\ref{eq_sk}) and
(\ref{eq_ask}) is the following
\begin{equation}
{\bf n}=\left(
\frac{2\lambda x}{r^2+\lambda^2},
\frac{2\lambda y}{r^2+\lambda^2},
\frac{r^2-\lambda^2}{r^2+\lambda^2}
\right),
\label{eq_hs}
\end{equation}
and
\begin{equation}
{\bf n}=\left(
\frac{2\lambda x}{r^2+\lambda^2},
-\frac{2\lambda y}{r^2+\lambda^2},
\frac{r^2-\lambda^2}{r^2+\lambda^2}
\right).
\label{eq_ahs}
\end{equation}

The topological charge $Q$ is given by
\begin{equation}
Q=\frac{1}{\pi} \int d^2 {\bf r}^2
\frac{
(\overline{\partial} \overline{w} )
(\partial w )
- 
(\partial \overline{w} )
(\overline{\partial} w )}
{(1+|w|^2)^2}.
\end{equation}
For the solutions Eqs.(\ref{eq_sk}) and (\ref{eq_ask}),
the topological charges are $Q=1$ and $Q=-1$, respectively.
These solutions are called skyrmion.
The energy of these skyrmion solution is 
$E=4\pi \rho_s$,
which is calculated from the following expression:
\begin{equation}
E=4\rho_s \int d^2 {\bf r} 
\frac{ (\overline{\partial} \overline{w} )
(\partial w )
+ 
(\partial \overline{w} )
(\overline{\partial} w )}
{(1+|w|^2)^2}.
\end{equation}
Note that there is no solution in the $Q=0$ sector.
In fact, solutions in this sector satisfy $\partial \omega=0$
and $\overline{\partial} \omega=0$,
and we obtain $\omega={\rm const.}$
Obviously such a solution does not satisfy the boundary conditions.

\begin{figure}[htbp]
\includegraphics[scale=0.5]{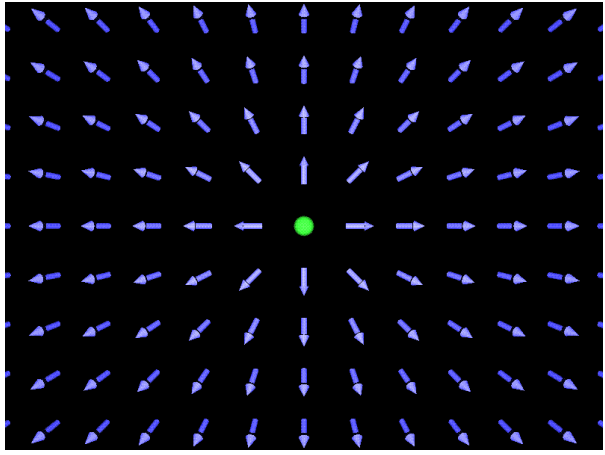}
\caption{The half-skyrmion spin texture. Arrows indicate the
staggered moment and the filled circle at the center represents the
Zhang-Rice singlet formed site.}
\label{fig_hs}
\end{figure}

\begin{figure}[htbp]
\includegraphics[scale=0.5]{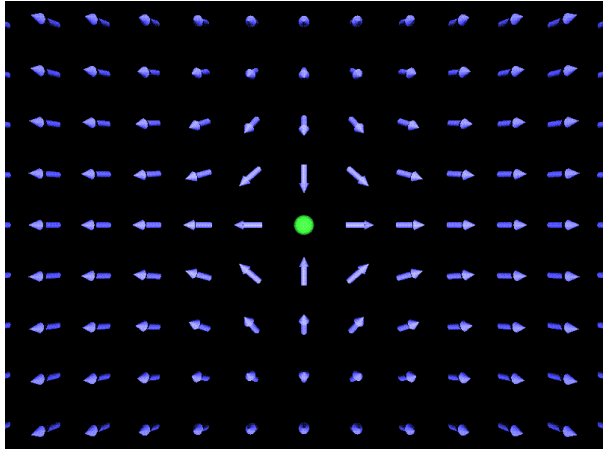}
\caption{The anti-half-skyrmion spin texture. Arrows indicate the
staggered moment and the filled circle at the center represents the
Zhang-Rice singlet formed site.}
\label{fig_ahs}
\end{figure}

Now we consider superposition of the uniform state and the
skyrmion state.
Unfortunately, the classical superposition of the two spin
configuration is not the solution of the field equation.
However, these solutions suggest that the resultant spin
configuration is chracterized by a topological charge $Q$ with
$0<|Q|<1$.
The value of $Q$ is determined by making use of the fact that N{\' e}el
ordering state is described by Bose-Einstein condensation of the
Schwinger bosons.\cite{BEC}
In order to examine the value of $Q$, we use the following
representation of $Q$ by the CP$^1$ gauge field $\alpha_{\mu}$,
\begin{equation}
Q=\int \frac{d^2 {\bf r}}{2\pi} 
\left( \partial_x \alpha_y - 
\partial_y \alpha_x \right).
\end{equation}
From this expression, we see that the spin configuration with $Q$
corresponds to the flux $2\pi Q$ in the condensate of the Schwinger
bosons.
Since the spin $1/2$ bosons $\zeta_{\sigma}(x)$ are confined in the
N{\' e}el state, all bosons are paired in the low-energy physics.
Because pairs of the bosons carry the gauge charge two,
the flux quantum is $\pi$ similar to the conventional BCS
superconductors.
Therefore, the flux value is not arbitrary and 
$Q$ must be in the form of $Q=n\pi$, with $n$ being an integer.
Meanwhile,
from the constraint $0<|Q|<1$, the flux associated with the spin
configuration satisfies $0<2\pi |Q|<2\pi$.
Thus, we conclude $2\pi |Q|=\pi$, or $|Q|=1/2$.\cite{WENG_ETAL}
Since the topological charge is one-half, we call this spin
configuration as half-skyrmion.
The energy associated with the half-skyrmion spin texture is $4\pi
\rho_s |Q|=2\pi \rho_s \equiv E_0$
because we can apply Eq.(\ref{eq_2}) outside the core region.
Due to the limitation of the effective theories, 
we cannot determine the core energy.
It would be determined from a calculation based on a microscopic
model.
The half-skyrmion solution is also discussed in the ferromagnetic
Heisenberg model\cite{SAXENA_DANDOLOFF} in the context of quantum Hall 
systems.
Since the value of $|Q|=1/2$ is obtained from the
calculation of the topological charge using Eqs.~(\ref{eq_hs}) and
(\ref{eq_ahs}) with excluding the core region $r<\lambda$,
we may use Eqs.~(\ref{eq_hs}) and (\ref{eq_ahs}) for the expressions
of the half-skyrmion and the anti-half-skyrmion spin textures,
respectively.
The half-skyrmion and anti-half-skyrmion spin textures are shown
schematically in Figs.~\ref{fig_hs} and \ref{fig_ahs}.

\section{Moving half-skyrmion solution}
\label{sec_moving}
In the last section, we have argued that the spin configuration around 
a static Zhang-Rice singlet is the half-skyrmion.
In this section, we construct the moving half-skyrmion solution from
the static half-skyrmion solution by making use of the Lorentz
invariance of the NL$\sigma$M and the CP$^1$ model.

The action (\ref{eq_nlsm}) is written in the Euclidean
space time.
The form in Minkowskii space time reads
\begin{equation}
S=\frac{\rho_s}{2} \int dt \int d^2 {\bf r}
\left[ 
\frac{1}{c_{\rm sw}^2} 
\left( \frac{\partial {\bf n}}{\partial t} \right)^2
- \left( \nabla {\bf n} \right)^2
\right].
\label{eq_nlsm_m}
\end{equation}
Apparently the action is invariant under Lorentz transformations with
$c_{\rm sw}$ being the speed of ``light.''
Let us consider a following Lorentz transformation,
\begin{equation}
x'=\frac{x-vt}{\sqrt{1-\left( v/c_{\rm sw} \right)^2}},
\label{eq_Lx}
\end{equation}
\begin{equation}
t'=\frac{t-\left(v/c_{\rm sw}^2 \right)t}
{\sqrt{1-\left( v/c_{\rm sw} \right)^2}}.
\label{eq_Lt}
\end{equation}
It is easy to check that the action is invariant under this
Lorentz transformation.

Now we apply the Lorentz transformation
(\ref{eq_Lx}) and (\ref{eq_Lt}) to the static half-skyrmion solution 
(\ref{eq_hs}):
\begin{equation}
{\bf n}'=
\left(
\frac{2\lambda x'}{x'^2+y^2+\lambda^2},
-\frac{2\lambda y}{x'^2+y^2+\lambda^2},
\frac{x'^2+y^2-\lambda^2}{x'^2+y^2+\lambda^2}
\right).
\label{eq_hsx}
\end{equation}
The excitation spectrum of the half-skyrmion is obtained by
calculating the energy-momentum tensors.
We write Eq.(\ref{eq_hsx}) as ${\bf n}'={\bf n}_s (x',y)$,
where $x'=\gamma (x-vt)$ with $\gamma=1/\sqrt{1-(v/c_{\rm sw})^2}$.
The energy-momentum tensors are given by
\begin{equation}
T^{\mu}_{\nu} = \rho_s \partial^{\mu} {\bf n} \partial_{\nu} {\bf n}
- \frac{\rho_s}{2} \partial^{\rho} {\bf n} \partial_{\rho} {\bf n}
\delta^{\mu}_{\nu},
\end{equation}
where $A^{\mu}B_{\mu} = A_0 B_0 - {\bf A} \cdot {\bf B}$.
The energy is 
\begin{eqnarray}
E&=&\int d^2 {\bf r} T^0_0 \nonumber \\
&=& \frac{\rho_s}{2} \int d^2 {\bf r} 
\left[ \frac{1}{c_{\rm sw}^2} \left( \partial_t {\bf n}' \right)^2
+ \left( \partial_x {\bf n}' \right)^2 
+ \left( \partial_y {\bf n}' \right)^2 \right] \nonumber \\
&=& \frac{\rho_s}{2} \int d^2 {\bf r} 
\left[ 
\left( \frac{\gamma^2  v^2}{c_{\rm sw}^2}+ \gamma^2 \right)
\left( \partial_{x'} {\bf n}_s \right)^2
+ \left( \partial_y {\bf n}_s \right)^2 \right]
\nonumber \\
&=&\gamma \rho_s \int dx' dy \left( \partial_{x'} {\bf n}_s \right)^2
\nonumber \\
&=&\gamma E_0,
\label{eq_E}
\end{eqnarray}
where we have used that 
$\partial_t {\bf n}' = -\gamma v \partial_{x'} {\bf
n}_s$, $\partial_x {\bf n}' = \gamma \partial_{x'} {\bf n}_s$,
$dx=\gamma dx'$, and 
$\left( \partial_{x'} {\bf n}_s \right)^2
=\left( \partial_y {\bf n}_s \right)^2$.
The $x$-component of the momentum is calculated as follows
\begin{eqnarray}
P_x &=& \frac{1}{c_{\rm sw}}
\int d^2 {\bf r} T^0_1 \nonumber \\
&=& \frac{\rho_s}{c_{\rm sw}^2} \int dx dy
\left( \partial_t {\bf n}_s \right) \cdot
\left( \partial_x {\bf n}_s \right) \nonumber \\
&=& -\beta \gamma E_0/c_{\rm sw},
\label{eq_Px}
\end{eqnarray}
\begin{eqnarray}
P_y &=& \frac{1}{c_{\rm sw}}
\int d^2 {\bf r} T^0_2 \nonumber \\
&=& \frac{\rho_s}{c_{\rm sw}^2} \int dx dy
\left( \partial_t {\bf n}_s \right) \cdot
\left( \partial_y {\bf n}_s \right) \nonumber \\
&=& 0.
\label{eq_Py}
\end{eqnarray}
From Eqs.(\ref{eq_E}), (\ref{eq_Px}), and (\ref{eq_Py}),
we find
\begin{equation}
E^2 = c_{\rm sw}^2  P_x^2 + E_0^2.
\end{equation}
By considering general Lorentz transformations, we find that 
the following relation holds:
\begin{equation}
E^2 = c_{\rm sw}^2  \left( P_x^2 + P_y^2 \right) + E_0^2.
\end{equation}
Therefore, the excitation spectrum of the half-skyrmion is given by
\begin{equation}
E_k = \pm \sqrt{c_{\rm sw}^2 k^2 + E_0^2}.
\label{eq_disp}
\end{equation}

Having obtained the relativistic dispersion (\ref{eq_disp}),
we consider the action of the half-skyrmion.
Since the half-skyrmion is a topological spin texture, one might
expect that Berry phases affect the statistice of the half-skyrmion as 
in the fractional quantum Hall systems.
In the fractional quantum Hall systems, 
the effective theories are characterized by topological field theory.
Berry phase effect determines the statistics of
quasiparticles.
In constrast, the system of single hole doepd antiferromagnet is not
characterized by topological field theory.
Indeed, a gauge field that describes the Berry phase effects is
massive due to Bose-Einstein condensation.
Therefore, the leading term of the gauge field is the mass term.
In such a situation, we do not expect that Berry phases play an
important role in determining the statistics of quasiparticles.
In the absence of the Berry phase effects, the statistice of the
half-skyrmion is fermion simply because the doped hole that obeys the
fermionic statistics sits at the core.

The statistics of the Zhang-Rice singlet is infered from its field
operator.
Since the d-orbital hole states constitute localized spin $1/2$
moments and one can choose either a fermionic description or a bosonic
description for the spins,
the statistics of the Zhang-Rice singlet is either fermion or boson.
However, now we are interested in the N{\' e}el ordering phase.
Bosonic descriptions, such as Schwinger boson mean field theory or
NL$\sigma$M, are suitable for the description of the N{\' e}el ordered
state.
In this case, the statistics of the Zhang-Rice singlet is fermion.

Fermion field obeying the relativistic excitation spectrum
(\ref{eq_disp}) is described by a Dirac fermion action.
The action of the half-skymion may be written as
\begin{equation}
{\cal L}=\sum_{s=\pm}
\overline{\psi}_s \left( \gamma_{\mu} \partial_{\mu}
+ m c_{\rm sw}^2 \right) \psi_s,
\label{eq_Dirac}
\end{equation}
with $\overline{\psi}_s = \psi_s^{\dagger} \gamma_0$
and $mc_{\rm sw}^2 = E_0$.
The index $s$ is for the sign of the topological charge.
That is, $s=+$ is for the half-skyrmion and $s=-$ is for the
anti-half-skyrmion.
The Dirac fermion $\psi$ has four components:
There are positive and negative energy states.
These states obey the dispersion (\ref{eq_disp}) with the origin
either at ${\bf k}_1 = \left( \pi/2,\pi/2 \right)$
or ${\bf k}_2 = \left( -\pi/2,\pi/2 \right)$.
This is suggested from the fact that the Schwinger bosons are gapless
at four zone centers $(\pm \pi/2, \pm \pi/2)$.
Because ${\bf Q}=(\pi,\pi)$ connects two points in the diagonal
directions, there are two independent points ${\bf k}_1$ nad ${\bf
k}_2$.
The $\gamma$ matrices are $4\times 4$ matrices and satisfy
$\gamma_{\mu} \gamma_{\nu} + 
\gamma_{\nu} \gamma_{\mu} =2\delta_{\mu \nu}$.

\section{Half-skyrmion on the lattice}
\label{sec_lattice}
The expression on the lattice is derived by discretizing 
the continuum Lagrangian Eq.(\ref{eq_Dirac}).
After Fourier transformations, we obtain
\begin{eqnarray}
{\cal L}&=&
\sum_{a=1,2} \sum_{\sigma}
\overline{\psi}_{a\sigma}(k) 
\left(
\begin{array}{cc}
mc_{\rm sw}^2+\partial_{\tau} & \sin k_x + i \sin k_y\\
-\sin k_x + i\sin k_y & mc_{\rm sw}^2-\partial_{\tau}\\
\end{array}
\right)
\psi_{a\sigma} (k).
\end{eqnarray}
By shifting the origin in the momentum space to either ${\bf k}_1$ or
${\bf k}_2$, we obtain
\begin{eqnarray}
{\cal L}
&=&\sum_{\sigma} 
\left[
\overline{\psi}_{1\sigma}(k)
\left(
\begin{array}{cc}
mc_{\rm sw}^2+\partial_{\tau} & \cos k_x + i \cos k_y\\
-\cos k_x + i\cos k_y & mc_{\rm sw}^2-\partial_{\tau}
\end{array}
\right) 
\psi_{1\sigma} (k). 
\right. \nonumber \\ & & \left.
+ \overline{\psi}_{2\sigma}(k)
\left(
\begin{array}{cc}
mc_{\rm sw}^2+\partial_{\tau} & -\cos k_x + i \cos k_y \\
\cos k_x + i\cos k_y & mc_{\rm sw}^2-\partial_{\tau}
\end{array}
\right)
\psi_{2\sigma} (k).
\right]
\end{eqnarray}
Diagonaliztion of the matrix leads to the following expression,
\begin{equation}
\epsilon_k^{\pm} = \pm \sqrt{c_{\rm sw}^2 
(\cos^2 k_x + \cos^2 k_y) + (mc_{\rm sw}^2)^2}.
\label{eq_disp_l}
\end{equation}
Note that there is no fermion doubling problem, which occurs when one
formulates a Dirac fermion on a lattice.\cite{KOGUT83}

The parameters $c_{\rm sw}$ and $mc_{\rm sw}^2=2\pi \rho_s$ are
determined from the values for the Heisenberg antiferromagnet.
We use $Z_c = 1.17$ and $Z_{\rho} = 0.72$ which are estimated from 
quantum Monte Carlo simulations\cite{QMC} and a series expansion
technique.\cite{SINGH}
Substituting these values into Eq.~(\ref{eq_disp_l}), 
we find that the band
width is $\sim 1.5$J and $m/J \sim 1.13$.
Meanwhile, experimentally estimated band width by Wells {\it et al}. 
is $\sim 2.2J$.\cite{WELLS_ETAL,SHEN_RMP}
This discrepancy would be associated with the deviation of the real
system from the NL$\sigma$M.
For the mass value, Ronning {\it et al}. evaluated it by using the
form $\epsilon_k^--mc_{\rm sw}^2$.
The result is $m/J \sim 1.3$.\cite{RONNING_ETALUN}
Some of this discrepancy might be associated with the mass
renormalization due to antiferromagnetic spin fluctuations.
This shall be discussed in Sec.\ref{sec_duality}.

\section{Effective theory for half-skyrmion}
\label{sec_duality}
In order to include antiferromagnetic spin fluctuation effects on the
half-skyrmion, we shall
derive the effective theory of the half-skyrmion.
We make use of a duality mapping\cite{FISHER_LEE} 
for that purpose.

Before the application of the duality mapping, we point out
that a half-skyrmion can be seen as a vortex in the CP$^1$ model.
As stated in Sec.\ref{sec_static}, the topological charge of the
half-skyrmion is represented by the gauge flux with respect to the
CP$^1$ gauge field $\alpha_{\mu}$.
In the CP$^1$ model, the vector ${\bf n}$ reads
\begin{equation}
{\bf n}
= \left(
\overline{\zeta}_{\uparrow} \zeta_{\downarrow}+
\overline{\zeta}_{\downarrow} \zeta_{\uparrow},
-i\left( 
\overline{\zeta}_{\uparrow} \zeta_{\downarrow}-
\overline{\zeta}_{\downarrow} \zeta_{\uparrow}
\right),
\overline{\zeta}_{\uparrow} \zeta_{\uparrow}-
\overline{\zeta}_{\downarrow} \zeta_{\downarrow}
\right).
\end{equation}
Since ${\bf n}$ is a unit vector, $\zeta$ and $\overline{\zeta}$
satisfy
$\overline{\zeta}_{\uparrow}\zeta_{\uparrow}+
\overline{\zeta}_{\downarrow}\zeta_{\downarrow}=1$.
In terms of $\zeta_{\sigma}$ and $\overline{\zeta}_{\sigma}$,
the half-skyrmion solution Eqs.(\ref{eq_hs})  and (\ref{eq_ahs}) has
the following form
\begin{eqnarray}
\zeta_{\uparrow}&=&\frac{r}{\sqrt{r^2+\lambda^2}}
\exp\left(\pm i\theta \right) \exp(i\chi), \\
\zeta_{\downarrow}&=&\frac{\lambda}{\sqrt{r^2+\lambda^2}}
\exp(i\chi),
\end{eqnarray}
where the minus sign is for the half-skyrmion and 
the plus sign is for the anti-half-skyrmion and
$\chi$ is a constant.
In general, the half-skyrmion solution is represented by
\begin{equation}
\left( \begin{array}{c}
\zeta_{\uparrow} \\
\zeta_{\downarrow}
\end{array}
\right)
= \left(
\begin{array}{cc}
u & -v^* \\
v & u^* \end{array}
\right) 
\left( \begin{array}{c}
\left( \lambda/\sqrt{r^2+\lambda^2} \right) \\
\left( r/\sqrt{r^2+\lambda^2} \right) 
\exp \left( \pm i\theta \right) \\
\end{array}
\right),
\end{equation}
where $u$ and $v$ are constant complex numbers and satisfy
$|u|^2+|v|^2=1$.
The matrix $\left(
\begin{array}{cc}
u & -v^* \\
v & u^* \end{array}
\right)$ is a global SU(2) transformation.
The boundary condition at infinity is transformed to 
${\bf n}\rightarrow \left( -uv^*-vu^*,-i(uv^*-vu^*),-|u|^2+|v|^2
\right).$

In order to see the relation between the half-skyrmion and a vortex,
we take $u=v=1/\sqrt{2}$.
In this case, the half-skyrmion solution have the following form
\begin{eqnarray}
\zeta_{\uparrow} &=& 
\frac{\lambda-r 
    \exp \left( {\pm i\theta} \right) }{\sqrt{2(r^2+\lambda^2)}},\\
\zeta_{\downarrow} &=& \frac{\lambda+r
  \exp \left( {\pm i\theta} \right)}{\sqrt{2(r^2+\lambda^2)}}.
\end{eqnarray}
This has a vortex form at $r\gg \lambda$:
\begin{equation}
\left( 
\begin{array}{c}
\zeta_{\uparrow} \\
\zeta_{\downarrow}
\end{array}
\right)
\sim \exp(\pm i\theta) 
\left(
\begin{array}{c}
-1/\sqrt{2}\\
1/\sqrt{2}
\end{array}
\right).
\label{eq_vortex}
\end{equation}
Thus, the half-skyrmion is seen as a vortex
introduced in the system.
Such a vortex is taken into account in the CP$^1$ model
as follows
\begin{equation}
\zeta_{\sigma} = \rho_{\sigma}^{1/2} \exp (i\phi)
=\rho_{\sigma}^{1/2} \exp (i\phi_0+i\phi_v).
\label{eq_amp_phase}
\end{equation}
Here $\phi_0$ describes coherent motion of the bosons and $\phi_v$
describes the vortex.
In case of Eq.(\ref{eq_vortex}), $\phi_v$ is 
\begin{equation}
\phi_v = \pm \tan^{-1} \frac{y-y_v}{x-x_v},
\end{equation}
with $(x_v,y_v)$ representing the coordinate of the vortex.

We rewrite the CP$^1$ model by substituting
Eq.(\ref{eq_amp_phase}) into Eq.(\ref{eq_cp1}):
\begin{equation}
S=\frac{1}{g} \int d^3 x
\left[ 
\sum_{\sigma} \frac{1}{4\rho_{\sigma}}
\left( \partial_{\mu} \rho_{\sigma} \right)^2
+ \left( \sum_{\sigma} \rho_{\sigma} \right)
\left( \partial_{\mu} \phi - \alpha_{\mu} \right)^2
\right].
\end{equation}
The amplitude fluctuations would be important only in the vicinity of
the core.
We focus on the outside of the core and assume a constant value for
$\rho_{\sigma}$: $\rho_0 \equiv \sum_{\sigma} \langle \rho_{\sigma}
\rangle$.
After introducing a Stratonovich-Hubbard field $J_{\mu}$, we obtain
\begin{equation}
S=\int d^3 x \left[ \frac{g}{4\rho_0} J_{\mu}^2 
-iJ_{\mu} \left( \partial_{\mu} \phi_0 +
\partial_{\mu} \phi_v
-\alpha_{\mu} \right)
\right].
\end{equation}
The field $J_{\mu}$ is associated with the spin current.
Integrating out $\phi_0$ leads to $\partial_{\mu} J_{\mu}=0$.
From this equation, we can represent $J_{\mu}$ in terms of a gauge
field
\begin{equation}
J_{\mu} = \frac{1}{2\pi} \epsilon_{\mu\nu\lambda} \partial_{\nu}
A_{\lambda}.
\label{eq_JA}
\end{equation}
Substituting this equation into the action, and integrating out
$\alpha_{\mu}$, we obtain 
\begin{equation}
S_d=S_A+S_{\rm int},
\end{equation}
where
\begin{equation}
S_A=\frac{1}{4e_A^2} \int d^3 x 
\left( \partial_{\mu} A_{\nu} - \partial_{\nu} A_{\mu} \right)^2,
\end{equation}
and
\begin{equation}
S_{\rm int} = -i\int d^3 x A_{\mu} j_{\mu}^v
\end{equation}
with
\begin{equation}
j_{\mu}^v = \frac{1}{2\pi} \epsilon_{\mu\nu\lambda}
\partial_{\nu} \partial_{\lambda} \phi_v.
\end{equation}
The value of the gauge charge $e_A$ depends on the action of the gauge 
field $\alpha_{\mu}$, which would be massive because of Bose-Einstein
condensation of the Schwinger bosons.
Here we treat $e_A$ as a parameter of the theory.

Since $J_{\mu}$ describes the spin current and $A_{\mu}$ is related to 
$J_{\mu}$ through Eq.~(\ref{eq_JA}), 
the gauge field $A_{\mu}$ is associated with spin
excitations, such as antiferromagnetic spin waves.
Indeed, $A_{\mu}$ is massless in the N{\' e}el ordered phase and the
velocity of the mode is $c_{\rm sw}$.

By taking into account the fact that the half-skyrmion is given by 
Eq.(\ref{eq_Dirac}), we obtain the following action for the
half-skyrmion
\begin{eqnarray}
S&=&\int d^3 x \left[
\sum_{\sigma} \overline{\psi}_{\sigma}
\left[ \gamma_{\mu} \left( \partial_{\mu}
-iq_{\sigma} A_{\mu} \right) + m \right] 
\psi_{\sigma} 
+ \frac{1}{4e_A^2}
\left( \partial_{\mu} A_{\nu}
-\partial_{\nu} A_{\mu}
\right)^2
\right],
\label{eq_Seff}
\end{eqnarray}
where $q_{\sigma}$ is an index for the sign of the topological charge.
Hereafter, we set $c_{\rm sw}=1$.

In order to study the effect of spin fluctuations on the
half-skyrmion, we formulate the theory on the square lattice.
The lattice form of (\ref{eq_Seff}) is given by
\begin{eqnarray}
{\cal L}&=&\sum_{a}\sum_{j} \sum_{\mu}
\left[
\frac12 \left(
\overline{\psi}_{ja} \gamma_{\mu} e^{-iq_{\sigma} \phi_{\mu} (j)}
\psi_{j+{\hat \mu},a} 
-\overline{\psi}_{j+{\hat \mu},a} \gamma_{\mu} e^{iq_{\sigma}
\phi_{\mu} (j)} \psi_{ja} \right) 
+ m \overline{\psi}_{ja} \psi_{ja} \right]
\nonumber \\
& & + \frac{1}{2e_A^2} \sum_j \sum_{\mu \nu}
\left\{ 1-\cos 
\left[ \phi_{\nu} (j+{\hat \mu}) - \phi_{\nu}(j)
-\phi_{\mu} (j+{\hat \nu}) + \phi_{\mu}(j) \right]
\right\}.
\end{eqnarray}
We expand $\exp(\pm i q_{\sigma} \phi_{\mu} (j))$ with respect to
$\phi_{\mu}(j)$.
The first order term ${\cal L}^{(1)}$ is 
\begin{eqnarray}
{\cal L}^{(1)} &=& -i \sum_{a}\sum_{\mu} \sum_{k,q}
q_{\sigma} \overline{\psi}_{k+q,a}
\gamma_{\mu} \phi_{\mu} (q) \psi_{k,a}
e^{-iq_{\mu}/2} 
\cos \left( k_{\mu} + \frac{q_{\mu}}{2} \right).
\end{eqnarray}
The second order term is
\begin{eqnarray}
{\cal L}^{(2)} &=& -\frac{i}{2}
\sum_{k,q,q'} \sum_{\mu} \sum_a
\overline{\psi}_{k+q+q',a} \gamma_{\mu}
\phi_{\mu} (q) \phi_{\mu} (q')
\psi_{k,a} e^{-i(q_{\mu}+q'_{\mu})/2}
\sin \left( k_{\mu} + \frac{q_{\mu}+q'_{\mu}}{2} \right).
\end{eqnarray}

We evaluate the self-energy $\Sigma_k^{(I)}$ 
that comes from the second order term of ${\cal L}^{(1)}$
and $\Sigma_k^{(II)}$ that comes from the first order term
of ${\cal L}^{(2)}$.
The self-energy $\Sigma_k^{(II)}$ is given by
\begin{equation}
\Sigma_k^{(II)}=\frac{i}{2\Omega} {\sum_q}' D_q \gamma_{\mu} 
\sin k_{\mu},
\end{equation}
with $D_q = 1/\left[ (i\omega_n)^2 - \omega_q^2 \right]$.
The mass renormalization due to this self-eneryg is
\begin{equation}
\frac{m}{1-\frac{1}{2\Omega} {\sum_q}' D_q}
= \frac{m}{1-0.123e_A^2}.
\end{equation}
The self-energy $\Sigma_k^{(I)}$ is given by
\begin{equation}
\Sigma_k^{(I)}=-\frac{2}{\Omega} {\sum_q}'
\gamma_{\mu} D_q G_{k+q} \gamma_{\mu} \cos^2 
\left( k_{\mu} + \frac{q_{\mu}}{2} \right).
\end{equation}
Substituting the explicit form of $G_k$,
\begin{equation}
G_k = \frac{1}{\sin^2 k_{\mu} + m^2}
\left( -i\gamma_{\mu} \sin k_{\mu} + m \right),
\end{equation}
we obtain
\begin{eqnarray}
\Sigma_k^{(I)}&=&-\frac{2e_A^2}{\Omega} {\sum_q}'
\frac{\left[ i\gamma_{\rho} \sin (k_{\rho} + q_{\rho}) 
+ mc^2 \right] \cos^2 \left(k_{\mu} + \frac{q_{\mu}}{2} \right)
-2i\gamma_{\mu} \sin (k_{\mu} + q_{\mu} )
\cos^2 \left( k_{\mu} + \frac{q_{\mu}}{2} \right)}
{\sin^2 q_{\nu} \left[ \sin^2 (k_{\nu}+q_{\nu}) + m^2\right]}
\nonumber \\
&=& -m e_A^2 \int \frac{d^3 q}{(2\pi)^3}
\frac{\sum_{\mu} \cos^2 \left( k_{\mu} + \frac{q_{\mu}}{2} \right)}
{\left( \sum_{\nu} \sin^2 q_{\nu} \right)
\left[ \sum_{\rho} \sin^2 (k_{\rho} + q_{\rho}) + m^2\right]}.
\end{eqnarray}
Since the dominant contribution comes from the region of 
$q_{\mu} \sim 0$ and $q_{\mu} \sim (\pi,\pi)$, 
we make an approximation for evaluating the
integral:
\begin{eqnarray}
\Sigma_k^{(I)}&\simeq &
-2m e_A^2 \int \frac{d^3q}{(2\pi)^3}
\frac{1}{q^2} \frac{\sum_{\nu} \cos^2 k_{\nu}}
{\sum_{\rho} \sin^2 k_{\rho} + m^2}
\nonumber \\
&=& -\frac{m e_A^2}{\pi} 
\frac{\sum_{\mu} \cos^2 k_{\mu}}
{\sum_{\nu} \sin^2 k_{\nu} + m^2}.
\end{eqnarray}
At the zone centers, the energy is shifted by $-2e_A^2/(\pi m)$.

\section{Discussion}
\label{sec_discussion}
On the description of the half-skyrmion spin texture formation,
it is essential that the Zhang-Rice singlet state has the form of
superposition of
the d-orbital spin up state and the d-orbital spin down state.
However, each of d-orbital spin states accompanies the oxygen
p-orbital hole state.
As argued by Aharony et al.,\cite{AHARONY_ETAL88} there is a
possibility of replacing the antiferromagnetic superexchange
interaction with a ferromagnetic interaction on the bond where the
doped hole occupies the oxygen p-orbital state.
In the picture of the Zhang-Rice singlet where a copper site hole spin 
forms a singlet state with the Wannier state of the symmetric
combination of the four oxygen p-orbital hole states surrounding the
copper site,
the interactions between the nearest neighbor spins could be replaced
with ferromagnetic interactions.
If this is the case, the skyrmion spin texture would be formed for the 
opposite spin state compared with the antiferromagnetic interaction
case.
However, obviously this does not affect the conclusion on the
half-skyrmion formation.
Another possibility on the effect of the oxygen hole state would be
related to the core structure of the half-skyrmion.
The discussion on the core structure is beyond the description of the
effective theories.
For the investigation of this effect, one would require a microscopic
model, such as the d-p model.

As argued in Sec.\ref{sec_duality}, the half-skyrmion spin texture can 
be seen as a vortex introduced in Bose-Einstein condensate of the
Schwinger bosons that describes the N{\' e}el ordering.
Generally, vortices disturb the phase coherence of condensate.
Rapid suppression of N{\' e}el ordering by hole doping could be understood
by this picture.\cite{TIMM_BENNEMANN}
We expect that this effective N{\' e}el order suppression is absent in the
electron doping case.
Our picture for the half-skyrmion formation is based on the fact that
the Zhang-Rice singlet has a form of superposition of the spin up
and spin down states.
This suggests that a picture for the difference between the hole
doping and the electron doping.
In the electron doping case, the doped electrons occupy copper sites.
The occupied copper site has spin zero.
However, the electronic state at this site does not have a form of
superposition of different spin states but is simply given by a doubly
occupied site.
Therefore, we do not expect the half-skyrmion formation for the
electron doped systems.
Similar arguments can be applied to Zn doping.
If we replace the copper with Zn, the electronic state at that site
does not have superposition of different spin states.
Since vortices are not induced by either the electron doping or the Zn 
doping, those dopings are not so effective to destroy the N{\' e}el
ordering as observed experimentally.
By constrast, if we replace a copper with Li, then a hole is
introduced.
That hole is believed to occupy an oxygen p-orbital state.
If we assume that the Zhang-Rice singlet formation between the hole
spin and a copper site spin, the half-skyrmion would be created.
This is consistent with the experiments that report that the critical
Li doping concentration for the destruction of the N{\' e}el ordering is
almost the same as that of the hole doping
concentration.\cite{Li_doping}
A skyrmion-like spin texture\cite{SKYRMIONS} 
formation for the Li-doping case was
discussed by Haas {\it et al}.\cite{HAAS_ETAL}

In Sec.~\ref{sec_lattice}, we have argued that the dispersion
(\ref{eq_disp_l}) is qualitatively in good agreement with the ARPES
result on the parent compounds.
A similarity between the dispersion (\ref{eq_disp_l}) without the mass 
term and the ARPES result was first pointed out by
Laughlin.\cite{LAUGHLIN97}
However, in the absence of the mass term cusp structures appear around 
the $(\pm \pi/2,\pm \pi/2)$ points which does not agree with the
experiment.
This discrepancy is corrected by including the mass term.
Recent estimation of the mass term \cite{RONNING_ETAL} reports a value 
which is close to the theoretical prediction as discussed in
Sec.~\ref{sec_lattice}.
Another point of view has been suggested based on a self-consistent
Born approximation \cite{KANE_ETAL} of the t-J model.
The analysis of the t-J model predicts a relatively flat dispersion
along the $(\pi,0)$ point to the $(0,\pi)$ poiint whose band width is
much smaller than the ARPES result.
This discrepancy is improved by taking into account next and third
nearest neighbor hopping terms.
Contrastingly, the dispersion (\ref{eq_disp_l}) predicts the same
dispersion along the $(0,0)$ point to the $(\pi,\pi)$ point and the
$(\pi,0)$ point to the $(0,\pi)$ point.
This seems consistent with the experiment by Ronning {\it et
al}.\cite{RONNING_ETAL}

The quite broad peaks observed by ARPES would be associated with the
coupling to some bosonic modes.
The question is what boson mode plays the major role for this
broadening.
A scenario based on an electron-phonon coupling has been proposed by
Mischenko and Nagaosa.\cite{MISCHENKO_NAGAOSA}
It is argued that the electron-phonon coupling is in the strong
coupling regime for the quasiparticles in the t-J model.
The quasiparticle dispersion in the t-J model is associated with the
antiferromagnetic spin wave effects.
Whereas in the half-skyrmion picture, the dispersion is given by the
soliton character of the half-skyrmion in the spin system.
The antiferromagnetic spin fluctuations can play some role for the
broadening of the spectrum.
This issue will be examined in the future publication.
This effect comes from the coupling between the half-skyrmion and the
gauge field that is associated with the antiferromagnetic spin
fluctuations as discussed in Sec.~\ref{sec_duality}.
Of course, this just suggests another possibility for the broadening
of the spectrum.


Ng argued \cite{NG99} that the vortex excitations 
in the Schwinger boson mean field theory 
correspond to the quasiparticles in the $\pi$-flux phase.
Similarities between the skyrmion-like spin texture
\cite{SKYRMIONS}
and the quasiparticle of the $\pi$-flux phase was pointed out by
Gooding \cite{GOODING}based on a numerical simulation.
These seem to be consistent with the half-skyrmion picture.
In the context of the spin-charge separation,\cite{ANDERSON} 
Baskaran argued \cite{BASKARAN} that the 
half-skyrmion can be seen as a deconfined spinon from the analysis of
the $n$-skyrmion solution of the NL$\sigma$M.

The half-skyrmion picture may be extended to the slightly doped
regime.
If we increase the number of holes, then the interaction between the
half-skyrmions becomes important.
A half-skyrmion lattice state might be possible in an appropriate
doping concentration.
We expect that a skyrmion-like spin texture is formed even in the
magnetically disordered phase.
Since the antiferromagnetic correlation length $\xi_{AF}$ is finite,
the topological charge that characterizes the spin texture is given by
\begin{equation}
Q=\frac{1}{2\pi} \int_{\lambda}^{\xi_{AF}} d^2 {\bf r}
\left( \partial_x \alpha_y -\partial_y \alpha_x \right).
\end{equation}
Probably the term ``topological'' is not appropriate in the disordered 
regime because the index $Q$ would be no longer quantized.
If doped holes accompany a spin-texture with non-zero $Q$, then we can 
apply a mechanism of d-wave superconductivity based on a skyrmion like 
spin texture.\cite{MORINARI02}

\section{Summary}
\label{sec_summary}
To summarize, it has been argued that the half-skyrmion spin texture
is created by doping a hole into the CuO$_2$ plane.
The picture is based on the Zhang-Rice singlet.
The Zhang-Rice singlet wave function has the form of superposition of
different d-orbital spin states.
The half-skyrmion formation is the result of superposition of the
skyrmion spin texture and the trivial uniform state associated with
each of d-orbital spin states.

The excitation spectrum of the half-skyrmion is qualitatively in good
agreement with ARPES experiments on the parent compounds.
Although the theory is based on the effective theories, such as the
NL$\sigma$M and the CP$_1$ model, 
estimation of the parameters in the dispersion gives us the values
close to experimentally obtained values.

Since the half-skyrmion picture is based on the superposition nature
of the Zhang-Rice singlet wave function,
the picture is not applicable to the electron doped systems
where the description of the doped electrons is not expected to have
that character.

Interesting extention of the half-skyrmion picture would be to examine 
whether the picture can be applicable to slightly doped regime where
the interaction between the half-skyrmions is not negligible.
It would be also interesting to examine whether skyrmion-like spin
texture is formed in the disordered spin state.
In the absence of the N{\' e}el ordering, the topological charge could
be arbitrary.
However, we expect that strong antiferromagnetic correlations can 
stabilize a skyrmion-like spin texture created by a doped hole with
non-vanishing topological charge.

\acknowledgments
I would like to thank G.~Baskaran, X.~Dai, N.~Nakai,
M.P.A.~Fisher, L.~Balents, K.~Shizuya, and A.~Tanaka for useful
discussions. 
This work was supported by Grant-in-Aid for Young Scientists(B) 
(17740253) and the 21st Century COE "Center for Diversity and
Universality in Physics" from the Ministry of Education, Culture,
Sports, Science and Technology (MEXT) of Japan.


\end{document}